\documentclass[%
 aip,
 amsmath,amssymb,
 reprint,%
]{revtex4-2}

\usepackage{graphicx}
\usepackage{dcolumn}
\usepackage{bm}
\usepackage[outercaption]{sidecap} 

\usepackage[utf8]{inputenc}
\usepackage[T1]{fontenc}
\usepackage{mathptmx}

\begin{document}

\preprint{AIP/123-QED}

\title{Growth modes and coupled morphological-compositional modulations in GaP$_{1-x}$N$_{x}$ layers grown on nominally (001)-oriented Si substrates}


\author{K. Ben~Saddik}
\affiliation{Electronics and Semiconductors Group (ElySe), Applied Physics Department, Universidad Autónoma de Madrid, 28049 Madrid, Spain}%

\author{S. Fernández-Garrido}
\email { Present address: ISOM and Materials Science Dept., Universidad Politécnica de Madrid, Avda. Complutense 30, 28040 Madrid, Spain }
\affiliation{Electronics and Semiconductors Group (ElySe), Applied Physics Department, Universidad Autónoma de Madrid, 28049 Madrid, Spain}%
\affiliation{\mbox{Centro de Microanálisis de Materiales, Universidad Autónoma de Madrid, 28049 Madrid, Spain}}
\affiliation{Instituto Nicolás Cabrera, Universidad Autónoma de Madrid, 28049 Madrid, Spain}

\author{R. Volkov}
\affiliation{National Research University of Electronic Technology MIET, Zelenograd, Russia}%


\author{J. Grandal}
\affiliation{ISOM, Universidad Politécnica de Madrid, Avda. Complutense 30, 28040 Madrid, Spain}%

\author{N. Borgardt}
\affiliation{National Research University of Electronic Technology MIET, Zelenograd, Russia}%

\author{B. J. García}%
\email{basilio.javier.garcia@uam.es}
\affiliation{Electronics and Semiconductors Group (ElySe), Applied Physics Department, Universidad Autónoma de Madrid, 28049 Madrid, Spain}%
\affiliation{\mbox{Centro de Microanálisis de Materiales, Universidad Autónoma de Madrid, 28049 Madrid, Spain}}
\affiliation{Instituto Nicolás Cabrera, Universidad Autónoma de Madrid, 28049 Madrid, Spain}

\date{\today}

\begin{abstract}
We comprehensively investigated the chemical beam epitaxy of GaP$_{1-x}$N$_{x}$ compounds to correlate the growth parameters with their properties when they are grown on nominally $(001)$-oriented Si substrates, as desired for the lattice-matched integration of optoelectronic devices with the standard Si technology. The growth mode as well as the chemical, morphological and structural properties of samples prepared using different growth temperatures and N precursor fluxes were analyzed by reflection high-energy electron diffraction, X-ray diffraction, Rutherford backscattering spectrometry, nuclear reaction analysis, energy dispersive X-ray spectroscopy, atomic force microscopy and transmission electron microscopy. Our results show that, up to $x\approx0.04$, it is possible to synthesize smooth and chemically homogeneous GaP$_{1-x}$N$_{x}$ layers with a high-structural quality in a two-dimensional (2D) fashion, namely, layer-by-layer. For a given N mole fraction, the layer-by-layer growth mode is favored by lowering the growth temperature while decreasing the N precursor flux. As the flux of the N precursor is increased at a given temperature to enhance N incorporation, the quality of the layers degrades upon exceeding a temperature-dependent threshold; above this threshold, the growing layer experiences a growth mode transition from 2D to 3D after reaching a critical thickness of a few nm. Following that transition, the morphology and the chemical composition become modulated along the $[110]$ direction with a period of several tens of nm. The surface morphology is then characterized by the formation of $\{113\}$-faceted wires, while the N concentration is enhanced at the troughs formed in between adjacent $(113)$ and $(\bar{1}\bar{1}3)$ facets. On the bases of this study, we conclude on the feasibility of fabricating homogeneous thick GaP$_{1-x}$N$_{x}$ layers lattice matched to Si ($x = 0.021$) or even with N content up to $x = 0.04$. The possibility of exceeding a N mole fraction of $0.04$ without inducing coupled morphological-compositional modulations has also been demonstrated when the layer thickness is kept below the critical value for the 2D--3D growth mode transition.
\end{abstract}

\maketitle

\section{\label{sec:level1}Introduction}
Highly-mismatched alloys (HMAs) are alloys in which host atoms are partially replaced by isoelectronic impurities of very different electronegativity and/or ionic radius.\cite{Walukiewicz2020} The differences are such that resulting alloys are characterized by a strong nonlinear dependence of their properties on the chemical composition. There are several examples where these materials are the subject of worldwide research as they are of practical interest to improve the efficiency and/or to reduce the production cost of a wide variety of optoelectronic devices.\cite{Walukiewicz2020} To cite a few, dilute-nitride and -antimonide III-V compounds are pursued to fabricate tandem solar cells and light-emitting devices,\cite{Harris2002,Olafsen2004,Gubanov2014,Mawst2015,Aho2017,Aho2018,Yamaguchi2021} Bi-containing III-V alloys have become a hot topic due to their potential for spintronics as well as for the fabrication of mid- to long-wavelength light sources and photodetectors,\cite{Marko2017,Bushell2019,Yoshimoto2019,Wang2019,Pacebutas2019} and Zn(Te,Se) compounds with dilute amounts of O are investigated as absorbing materials for intermediate band solar cells.\cite{Wang2009,Wang2011,Wu2013,Tanaka2018} Regardless of the particular material system, the enormous potential of widely tuning the electronic and optical properties of conventional compounds by adding dilute amounts of impurities is counterbalanced by the challenging synthesis of HMAs, as the combination of highly dissimilar elements typically results in alloy inhomogeneities and the creation of both point and structural defects.\cite{Suemune2000,Harris2002,Tang2016,Almosni2016,Tait2018,Luna2019,Stringfellow2021}

Among HMAs, dilute-nitride (Ga,In)(As,P,N) compounds are of major technological interest since these compounds are one of the very few choices to pseudomorphically integrate III-V solar cells, light-emitters and photodetectors on Si.\cite{Nagarajan2013,Rolland2014,Jain2014,DaSilva2015,Almosni2016,Yamane2017,Kharel2018,Yamaguchi2021} Regarding photovoltaics, this material system is furthermore particularly well suited to realize double- and triple-junction solar cells with almost perfect band gap combinations to maximize the photoconversion efficiency when grown on top of a Si cell (1.1/1.7~eV and 1.1/1.5/2.0~eV, respectively).\cite{Yamaguchi2021} In the specific case of the ternary GaP$_{1-x}$N$_{x}$, it is lattice matched to Si for $x = 0.021$ with a direct band gap energy of about $1.96$~eV at room temperature. Hence, the band gap energy of this ternary closely matches the ideal one for the topmost absorber in a triple-junction solar cell when combined with Si. Despite of the enormous potential of this material, its growth on nominally $(001)$-oriented Si wafers, as required for the lattice-matched integration of GaP$_{1-x}$N$_{x}$-based solar cells with the standard Si technology, was barely explored. GaP$_{1-x}$N$_{x}$ layers are instead commonly grown on highly misoriented ($4-6$~$^{\circ}$) Si substrates to avoid anti-phase domains,\cite{Furukawa2002,Utsumi2003,Utsumi2006,Yonezu2008,Nagarajan2013,Lazarenko2017,Kryzhanovskaya2017,Bolshakov2019,BenSaddik2019} a well-known problem that undermined the integration of III-V devices on CMOS compatible Si wafers.\cite{Yonezu2002,Li2017,Cornet2020} Only very recently, some of the present authors reported on the growth by chemical beam epitaxy (CBE) of GaP$_{1-x}$N$_{x}$ layers on nominally $(001)$-oriented Si substrates.\cite{Saddik2021} That work was addressed to study growth conditions, such as the dependence of the N content on the growth parameters founding the existence of two distinct growth regimes. On the one hand, there is a single-phase growth regime that results in flat layers with a homogeneous chemical composition on a macroscopic scale, as determined by high-resolution X-ray diffraction (HRXRD) reciprocal space maps (RSMs). On the other hand, there is a phase-separation growth regime that results in rough layers with two different chemical compositions, according to HRXRD measurements. This previous study also concludes on the feasibility of obtaining smooth and single-phase GaP$_{1-x}$N$_{x}$ layers on Si$(001)$ up to \textit{x} values of about $0.04$. Present work is adressed to a deeper post-growth characterization focusing on the study of the growth modes, the structural quality, and the microscopic distribution of N in single-phase and phase-separated layers.

In this work, we examine the growth mode as well as the chemical, morphological and structural properties of GaP$_{1-x}$N$_{x}$ layers grown by CBE on nominally $(001)$-oriented Si substrates as a function of the growth parameters using reflection high-energy electron diffraction (RHEED), HRXRD, Rutherford backscattering spectrometry (RBS), nuclear reaction analysis (NRA), energy dispersive X-ray spectroscopy (EDX), atomic force microscopy (AFM), and transmission electron microscopy (TEM). Such a comprehensive study allows us to demonstrate that: (i) up to \textit{x} values of approximately $0.04$, it is possible to grow GaP$_{1-x}$N$_{x}$ in a layer-by-layer fashion that results in epitaxial films with a smooth surface morphology, a high structural quality and a homogeneous chemical composition, and (ii) as the N content is increased, by either increasing the flux of the N precursor or decreasing the growth temperature, the GaP$_{1-x}$N$_{x}$ quality eventually degrades resulting in defective layers with a corrugated surface morphology coupled to a lateral composition modulation. Nevertheless, in the latter case, both the growth front roughing and the lateral compositional modulations only occur after exceeding a certain critical thickness, which leaves open the possibility of incorporating N mole fractions beyond $x = 0.04$ without inducing phase separation in layers with a thickness of only a few nm. 

\section{Experimental}

\begin{figure}
\includegraphics*[width=0.5\textwidth]{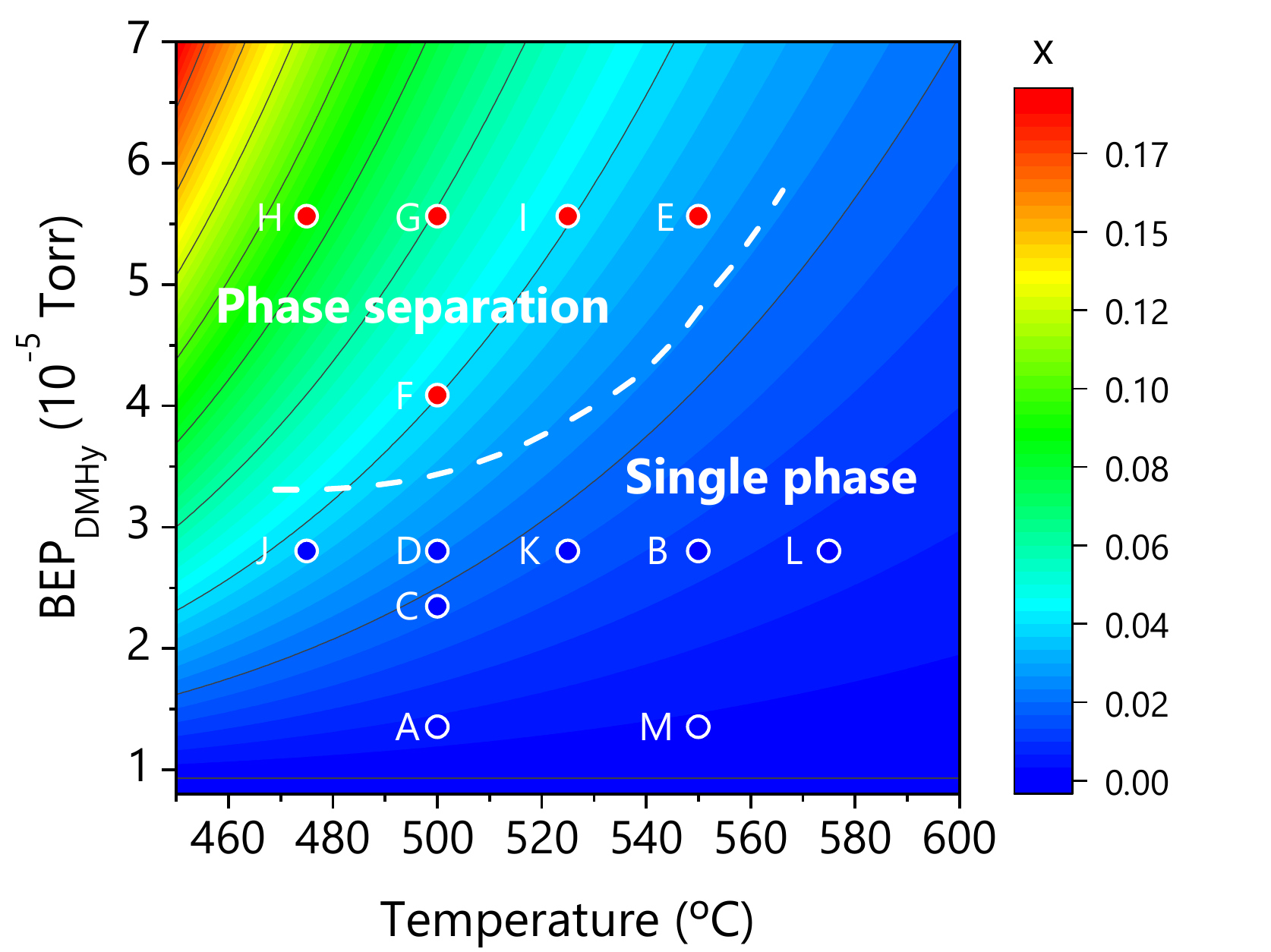}
\caption{Growth conditions of the samples studied in this work (solid symbols). The growth conditions are depicted over the growth diagram reported in Ref.\citenum{Saddik2021} for the synthesis of GaP$_{1-x}$N$_{x}$ by CBE. The diagram illustrates the dependence of the N mole fraction \textit{x} on the growth temperature and the BEP of DMHy for constant supplies of TBP ($BEP_{\mathrm{TBP}} = 1.2 \times 10^{-5}$~Torr) and TEGa ($BEP_{\mathrm{TEGa}} = 0.3 \times 10^{-5}$~Torr). The N mole fraction is displayed as a contour plot with a linear scale. The dashed line indicates the boundary between growth conditions resulting in chemically single-phase and phase-separated layers. Blue and red solid symbols indicate single-phase and phase-separated samples, respectively, as determined by HRXRD.}
\label{Figure1}
\end{figure}

\begin{table*}
  \caption{Type of sample (single-phase or phase-separated according to HRXRD measurements), rms roughness as determined by the analysis of $1\times1$~$\mu$m$^{2}$ atomic force micrographs, and mean N mole fraction \textit{x} derived from HRXRD, RBS-NRA and SEM-EDX and TEM-EDX for several GaP$_{1-x}$N$_{x}$ layers. For the phase-separated samples, we indicate in brackets the two main compositions derived from the combined analysis of of RSMs around the $004$ and $115$ Bragg reflections. The groups of single-phase and phase-separated samples are listed according to their mean N mole fraction.}
  \label{tbl:Table1}
  \begin{tabular}{lllllll}
    \hline
    Sample   &   Type   &  rms~(nm) & x$_{HRXRD}$   &   x$_{RBS-NRA}$   &   x$_{SEM-EDX}$ &   x$_{TEM-EDX}$(interlayer/layer)\\
    \hline
    A   & single phase & 0.3 & 0.007   & -- & $0.009\pm0.002$ & --\\
    B   & single phase & 0.3 & 0.011   & -- & -- & --\\
    C   & single phase & 0.4 & 0.016   & 0.015 & $0.021\pm0.003$ & --\\
    D   & single phase & 0.3 & 0.024   & -- & $0.023\pm0.003$ & $0.022\pm0.006$ \\
  	\hline    
    E   & phase separated & -- & 0.028 (0.026/0.031)  & 0.027 & -- & -- \\
    F   & phase separated & -- & 0.045 (0.036/0.056)   & -- & $0.036\pm0.005$ &$0.038\pm0.01$ (0.052/0.027) \\
    G   & phase separated & 1.9 & 0.06 (0.052/0.068)  & -- & $0.061\pm0.007$ & $0.058\pm0.01$ (0.095/0.05) \\
    H   & phase separated & -- & 0.08 (0.072/0.089)   & 0.1 & $0.091\pm0.009$ & -- \\
    \hline
  \end{tabular}
\end{table*}

All samples were grown by CBE on $2\times2$~cm$^{2}$ GaP-on-Si$(001)$ substrates diced from a 12" wafer purchased from\break $\mathrm{NAsP_{III/V}}$. Despite of the nominally exact $(001)$ orientation of the wafer, it exhibits a slight miscut of about $0.3^{\circ}$ towards one of the $\left<110\right>$ directions. The GaP layer, free of stacking faults and twins, is 25~nm thick and possesses a smooth surface without anti-phase domains. As gas sources, we used triethylgallium (TEGa), tertiarybutylphosphine (TBP) and 1,1-dimethylhydrazine (DMHy) for Ga, P and N, respectively. Low-temperature (120~$^{\circ}$C) gas injectors were used for both TEGa and DMHy, while a high-temperature one (820~$^{\circ}$C) was employed for TBP in order to increase the cracking efficiency. The beam equivalent pressure (BEP) of the impinging fluxes was assessed by an ion gauge that can be inserted to replace the substrate heater at the growth positon. As described in Ref.~\citenum{Saddik2021}, for Ga and P, the BEP values of the gas precursors were correlated with their corresponding elemental effective fluxes in equivalent growth rate units of monolayers per second (ML/s). Notice that, on Si(001), 1 ML corresponds to $2/a_{\mathrm{Si}}^{2}$, where $a_{\mathrm{Si}}$ is the Si lattice constant, i.\,e., 1~ML~$=6.78\times 10^{14}$~ atoms/cm$^{2}$. The growth temperature was measured using an optical pyrometer and the epitaxial processes were monitored in situ by RHEED using a 15~keV electron gun. Further technical details about the CBE sytem, a Riber CBE32, can be found elsewhere.\cite{AitLhouss1994} 

Prior to the growth of GaP$_{1-x}$N$_{x}$, the substrates were outgassed under TBP supply inside the growth chamber at 610~$^{\circ}$C until observing the appearance of a single domain $(2\times4)$ surface reconstruction. Afterwards, we grew a 15~nm thick GaP buffer layer at 580~$^{\circ}$C with $BEP_{\mathrm{TBP}}=1.2\times10^{-5}$~Torr and $BEP_{\mathrm{TEGa}}=0.3\times10^{-5}$~Torr, values that correspond to fluxes of 0.5 and 0.22~ML/s, respectively. After the growth of the GaP buffer layer, GaP$_{1-x}$N$_{x}$ layers were grown for 1~hour with a nominal Ga-limited growth rate of $0.22$~ML/s, which results in a layer thickness of $\approx200$~nm. All GaP$_{1-x}$N$_{x}$ layers were grown using the same fluxes of TBP and TEGa as for the growth of GaP. The different samples investigated here differ from each other in the flux of DMHy and/or the growth temperature used to grow the GaP$_{1-x}$N$_{x}$ layer. Figure~\ref{Figure1} summarizes their specific growth conditions  by positioning them in the growth diagram recently established by Ben Saddik et al.\cite{Saddik2021} The diagram illustrates both the dependence of the mean N mole fraction on the growth temperature and the BEP of DMHy, and the boundary between those growth conditions resulting in chemically single-phase and phase-separated GaP$_{1-x}$N$_{x}$ layers [according to HRXRD-RSMs]. The N composition of selected the samples, extracted from HRXRD RSMs around the $004$ and $115$ Bragg reflections, are collected in Table~\ref{tbl:Table1}. Additional information about the HRXRD measurements can be found in Ref.\citenum{Saddik2021}.

The morphology and surface roughness of the samples were investigated by AFM using either a Veeco Multimode or a Brucker Dimension Icon microscopes. In both cases, AFM micrographs were acquired in tapping mode. The micrographs were indifferently treated and analyzed with NanoScope or WSxM software.\cite{Horcas2007}. A Zeiss Ultra 55 SEM equipped with an EDX system was   used to assess the chemical composition of the GaP$_{1-x}$N$_{x}$ layers on a microscopic scale. The EDX spectra were acquired over an area of $0.01$~mm$^{2}$ using an electron aceleration voltage of 4~kV. The spatial resolution of the SEM-EDX measurements, estimated by Monte Carlo simulations using the software CASINO\cite{Drouin2007}, is about $30$~nm. 

RBS and NRA were used to independently determine the total amount of N incorporated into the samples. These measurements were carried out in a 5~MV High Voltage Engineering Europa tandem accelerator, equipped with two different detectors, using a collimated beam of $^{4}$He$^{+}$ ions.\cite{RedondoCubero2021a} RBS spectra were primarily acquired using a $2$~MeV $^{4}$He$^{+}$ ion beam. NRA spectra were, however, always acquired using a $3.72$~MeV $^{4}$He$^{+}$ ion beam in order to induce the resonant $^{14}$N($\alpha$,p)O$^{17}$ nuclear reaction. For the NRA measurements, one detector was covered by an 18~$\mu$m thick mylar foil to separate the proton signal from the large background caused by backscattered $\alpha$ particles. A second uncovered detector was used to simultaneously acquire the RBS spectrum at $3.72$~MeV in order to measure the particle dose in the NRA experiment. The P and N mole fractions in the GaP$_{1-x}$N$_{x}$ layers were obtained by the consistent fit of three different spectra, namely, RBS at $2$ and $3.72$~MeV, and NRA at $3.72$~MeV. The fits to the experimental RBS and NRA spectra were performed using the software SIMNRA.\cite{Mayer1999}

\begin{figure*}
\centering
\includegraphics*[width=0.75\textwidth]{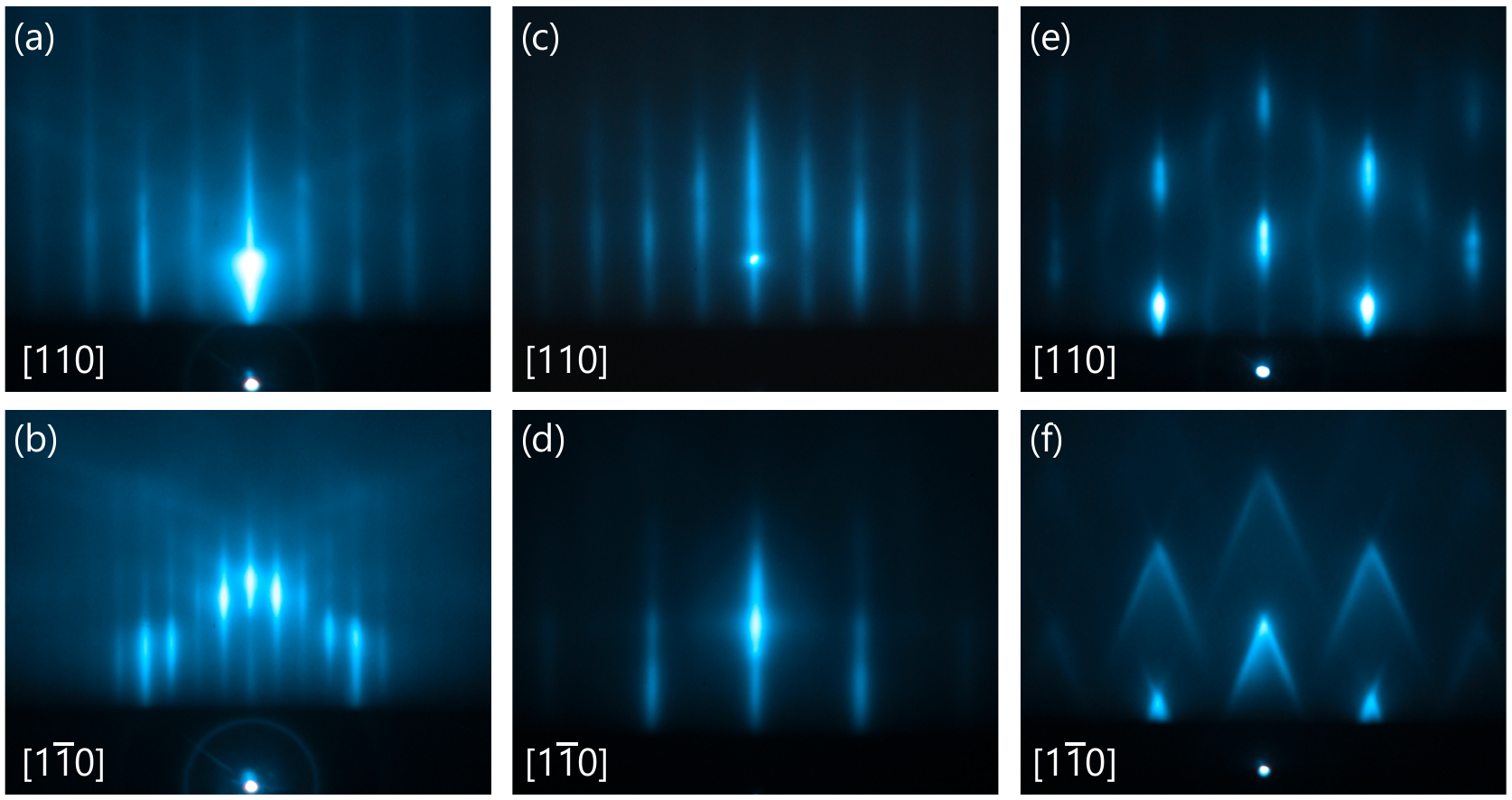}
  \caption{Exemplary RHEED patterns observed along the $[110]$ (top panels) and $[1\bar{1}0]$ (bottom panels) azimuths after the growth of: (a) and (b) the GaP buffer layer, (c) and (d) single-phase GaP$_{1-x}$N$_{x}$ layers (sample C), and (e) and (f) phase-separated GaP$_{1-x}$N$_{x}$ layers (sample I).}
  \label{Figure2}
\end{figure*}

TEM was utilized to study the structure and chemical composition of several samples (C, D, F and G in Fig.~\ref{Figure1}). The preparation of the specimens, which are thin films with a thickness enough for accelerated electron transmission, was performed using the standard \textit{in situ} lift-out technique\cite{Giannuzzi2005} on a FEI Helios NanoLab 650 DualBeam workstation equipped with a Kleindick MM3A-EM micromanipulator and a gas injection system for the local deposition of Pt and an amorphous C mix. Cross-sectional specimens were formed parallel to Si$(1\bar{1}0)$ substrate plane. For sample~G, an additional plan-view specimen was cut out parallel to the Si$(001)$ substrate surface from the upper half of the GaP$_{1-x}$N$_{x}$ layer. Two regions with thicknesses of about 80 and 10~nm were created for each specimen to study the chemical composition and the crystal structure, respectively. All TEM studies were carried out using a Titan Themis 200 microscope operated at 200~kV and equipped by a Cs image-corrector, a Fischione M3000 high-angular annular dark field detector (HAADF), and a Super-X EDX detector. Brightfield (BF) TEM mode was used to identify lattice defects, the presence of which caused a specific diffraction contrast in the images. Scanning TEM (STEM) images, acquired with low magnification by a HAADF detector, allowed to distinguish regions with different material densities, while high-resolution TEM and STEM (HRTEM and HRSTEM) micrographs provided a way of examining the specimen atomic structure. EDX spectroscopy was applied to the determination of the local specimen composition as well as to acquire compositional maps. An exact zone axis orientation of the specimens was used for obtaining the electron micrographs, while the specimens were deviated from such an orientation by a few degrees to avoid artefacts caused by electron channelling when acquiring chemical compositional maps. Quantitative EDX analysis was perfermed by Cliff-Lorimer's method with X-ray absorption correction, where the specimens thicknesses were determined by comparison of experimental and modeled convergent electron beam diffraction (CBED) patterns.

\section{Results}
\subsection{Growth modes and surface morphology}
\begin{figure*}
\centering
\includegraphics*[width=0.8\textwidth]{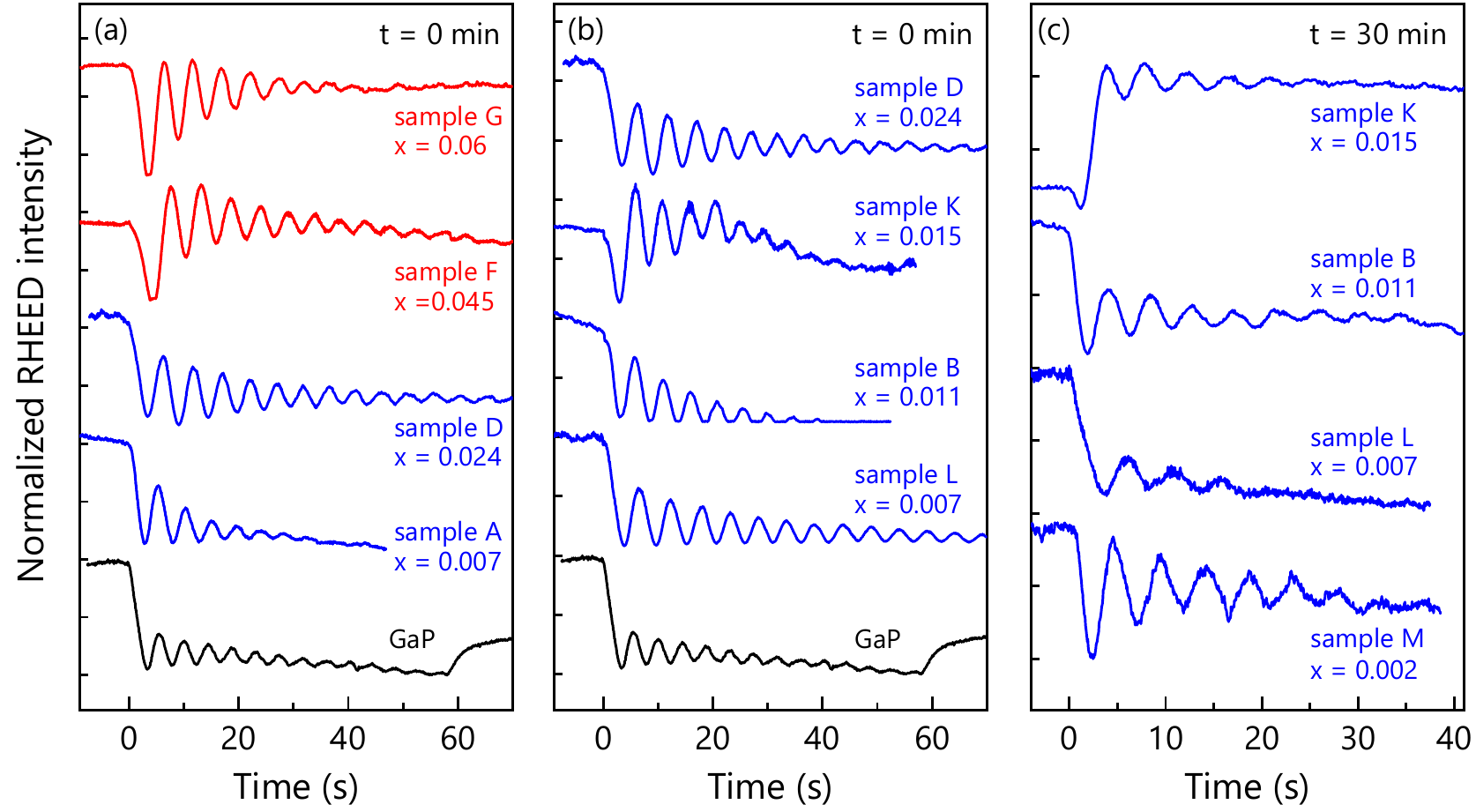}
\caption{(a and b) Specular $(0,0)$ RHEED spot intensity evolution at the beginning of the growth of GaP$_{1-x}$N layers grown with increasing N mole fractions by (a) rising $BEP_{\mathrm{DMHy}}$ at 500~$^{\circ}$C and (b) by decreasing the growth temperature with $BEP_{\mathrm{DMHy}}=2.8\times 10^{-5}$~Torr. We have also included for comparison exemplary RHEED intensity oscillations recorded during the growth of a GaP buffer layer. (c) Specular $(0,0)$ RHEED spot intensity evolution for selected samples upon interrupting and resuming the growth of GaP$_{1-x}$N$_{x}$ at 30~min. Blue and red data correspond to chemically single-phase and phase separated samples, respectively, while black ones correspond to GaP.}
\label{Figure3}
\end{figure*}
The growth mode of the GaP$_{1-x}$N$_{x}$ layers was analyzed \textit{in situ} by RHEED. The study revealed a strong correlation between the diffraction patterns and the chemical homogeneity of the samples. Figure~\ref{Figure2} presents representative RHEED patterns, along the $[110]$ and $[1\bar{1}0]$ azimuths, recorded immediately after the growth of the GaP buffer layer, a single-phase sample (sample~C), and a phase-separated sample (sample~I). For all GaP buffer layers, we observed a $(2\times 4)$ reconstruction with long streaks along both azimuths, with an intense specular spot superimpossed to the central streak, indicating a planarized surface, as shown in Figs.~\ref{Figure2}(a) and \ref{Figure2}(b). In the case of single-phase GaP$_{1-x}$N$_{x}$ layers, the RHEED patterns were also streaky, as those presented in Figs.~\ref{Figure2}(c) and \ref{Figure2}(d) for sample~C. We observed that during the growth of single-phase GaP$_{1-x}$N$_{x}$ layers, the surface reconstruction changes from $(2\times 4)$ to $(2\times 1)$ when the growth temperature is reduced below $\approx 530$~$^{\circ}$C (not shown here). On the contrary, the RHEED patterns of phase-separated GaP$_{1-x}$N$_{x}$ layers are remarkably different. For this type of samples, upon initiating the growth of the GaP$_{1-x}$N$_{x}$ layer, the diffraction pattern evolves from a streaky one to a a 3D pattern consisting in a highly modulated twofold reconstruction along the $[110]$ azimuth, with the appearance of pairs of streaks, known as chevrons, along the orthogonal $[1\bar{1}0]$ direction, as shown in Figs.~\ref{Figure2}(e) and \ref{Figure2}(f) for sample~I (azimuths are assigned according to Ref.\citenum{Grassman2009}). The observation of chevrons reveals the formation of inclined $\{113\}$ and $\{\bar{11}3\}$ facets, as deduced from the chevron angles measured on raw RHEED images, $(46\pm3)^{\circ}$, and further confirmed by cross-sectional transmission electron micrographs acquired along the $[1\bar{1}0]$ direction (see section~\ref{TEM section}). 
\begin{figure*}
\centering
\includegraphics*[width=\textwidth]{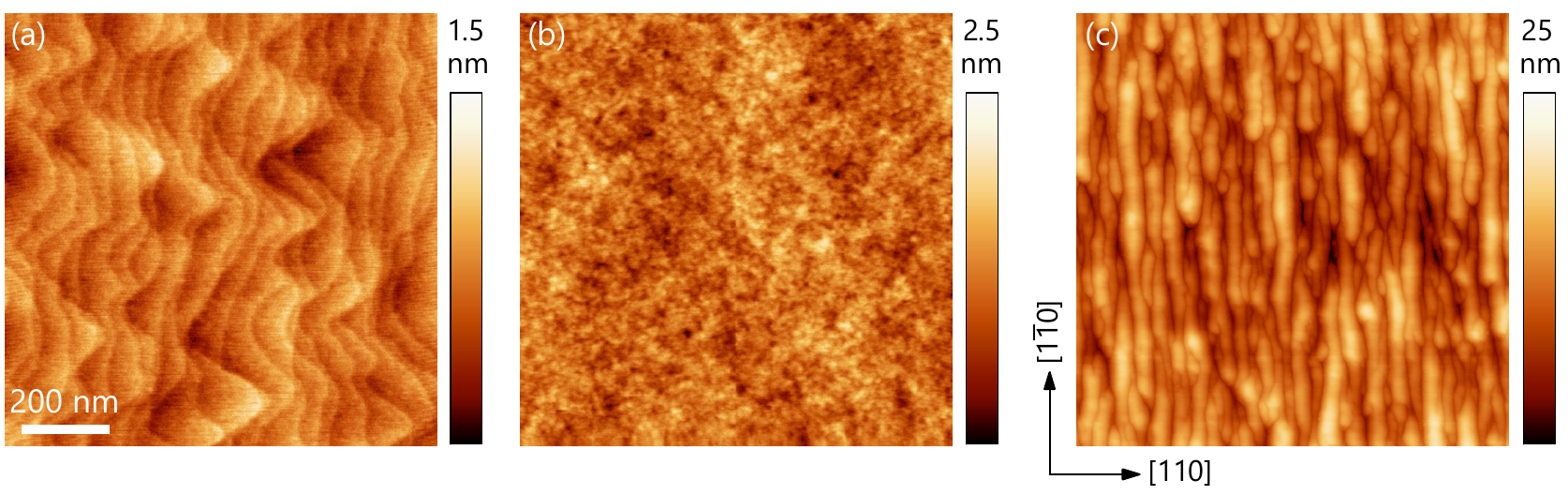}
  \caption{(a) Representative $1~\mu$m~$\times1~\mu$m atomic force micrographs of (a) the as-received GaP-on-Si substrate, (b) a chemically single-phase GaP$_{1-x}$N$_{x}$ layer (sample J), and (c) a chemically phase-separated GaP$_{1-x}$N$_{x}$ layer (sample G).}
  \label{Figure4}
\end{figure*}

As follows from the RHEED patterns shown in Fig.~\ref{Figure2}, while the growth mode of single-phase GaP$_{1-x}$N$_{x}$ layers is either step-flow or layer-by-layer (both growth modes lead to streaky RHEED patterns), the growth mode of phase-separated samples is 3D, at least upon exceeding a certain critical thickness. To gain further insights into the growth modes of the samples, we analyzed the temporal evolution of the RHEED intensity during the growth of the GaP$_{1-x}$N$_{x}$ layers. Figures~\ref{Figure3}(a) and \ref{Figure3}(b) show the RHEED intensity transients measured on the specular spot at the beginning of the growth of single-phase and phase-separated GaP$_{1-x}$N$_{x}$ layers (shown in blue and red colors, respectively). These figures summarize the results for two series of samples where N incorporation was enhanced in different manners, namely, by increasing the flux of DMHy [Fig.~\ref{Figure3}(a)], or by lowering the substrate temperature [Fig.~\ref{Figure3}(b)], while keeping constant the remaining growth parameters. For comparison, we included in both figures a RHEED intensity transient acquired during the growth of a GaP buffer layer. The detection of RHEED intensity oscillations during the growth of GaP reveals a layer-by-layer growth mode, also known as Frank-van der Merwe (FM) growth. With respect to the GaP$_{1-x}$N$_{x}$ layers, regardless of how the N content is increased, we observed strong RHEED intensity oscillations at the beginning of the growth, even in the case of phase-separated samples. Therefore, initially, GaP$_{1-x}$N$_{x}$ grows layer-by-layer. Since for phase-separated samples we recorded up to $>10$ oscillations, the layer-by-layer growth mode was sustained for a minimum thickness of 2--3~nm prior to the growth mode transition responsible for the 3D patterns shown in Figs.~\ref{Figure2}(e) and \ref{Figure2}(f). The specific critical thickness for the 2D-3D transition might, however, depend on both the N content and the specific growth conditions employed to incorporate the targeted amount of N.

To investigate whether the layer-by-layer growth mode is maintained during the growth of single-phase GaP$_{1-x}$N$_{x}$ layers, we made a growth interruption after 30~min by closing the supply of TEGa for some seconds. Then, upon resuming the growth, we registered the evolution of the specular spot intensity, observing clear RHEED intensity oscillations, as shown in Fig.~\ref{Figure3}(c) for several samples with various N mole fractions. We thus conclude that the layer-by-layer growth mode is kept during the growth process in that set of samples, as one could expect according to the streaky RHEED patterns observed after finishing the growth [Figs.~\ref{Figure2}(c) and \ref{Figure2}(d)]. 


To corroborate the results derived from the analysis of the samples by RHEED as well as to properly characterize the surface morphology of our GaP$_{1-x}$N$_{x}$ layers, several samples were examined by AFM. Figures~\ref{Figure4}(a), \ref{Figure4}(b) and \ref{Figure4}(c) present exemplary $1~\mu$m~$\times~1~\mu$m atomic force micrographs of a GaP/Si(001) substrate, a single-phase GaP$_{1-x}$N$_{x}$ layer, and a phase-separated GaP$_{1-x}$N$_{x}$ layer, respectively. The surface morphology of the as-received GaP-on-Si substrate is extremely smooth with root-mean-square (rms) roughness values below $0.2$~nm. The surface is further characterized by the presence of monolayer-height atomic steps separated by $50-60$~nm wide terraces [see Fig.~\ref{Figure4}(a)], values consitent with the measured $0.3^{\circ}$ miscut of the Si substrate.

As inferred from the micrograph of the single-phase sample [Fig~\ref{Figure4}(b)], the atomic steps of the substrate smear out until disappearing after the layer-by-layer growth of GaP and GaP$_{1-x}$N$_{x}$ in the CBE system, due either to the lack of surface annealing after growth, or to a a reduced adatom surface mobility as compared to that during the epitaxial process of GaP on Si by MOCVD. The resulting surface morphology of the single-phase layers is isotropic and very smooth. As summarized in Table~\ref{tbl:Table1}, the root mean square (rms) roughness values of these samples are $0.3-0.4$~nm. The roughness is thus only marginally larger than that of the GaP/Si$(001)$ substrate and consistent with the layer-by-layer growth mode deduced from the RHEED intensity transients shown in Fig.~\ref{Figure3}.

The surface morphology of the phase-separated GaP$_{1-x}$N$_{x}$ layers is totally different [see Fig.~\ref{Figure4}(c)]. These samples are not only rougher, with rms roughness values close to $2$~nm, but also exhibit a remarkably anisotropic surface morphology. As shown in Fig.~\ref{Figure4}(c), the surface becomes corrugated due to the formation of elongated 3D-islands, hereafter referred to as wires, parallel to the $[1\bar{1}0]$ direction with a periodicity of several tens of nm (about $40$~nm in the particular case of sample G). According to the \textit{in situ} analysis of the surface during growth by RHEED [Fig.~\ref{Figure2}(f)] and the \textit{ex situ} TEM study presented in section~\ref{TEM section}, these wires are enclosed by inclined $\{113\}$ facets. This peculiar surface morphology is analogous to those reported for compressively-strained GaAs\cite{Ohlsson1998} and tensilely-strained GaAs$_{1-x}$N$_{x}$\cite{Suemune2000} layers grown on GaP$(001)$ and GaAs$(001)$ substrates, respectively. In the case of GaAs on GaP, the formation of the wires was attributed to the inequivalent properties of type A and B facets, which result in the elastic relaxation of 3D-GaAs islands along the direction perpendicular to the wires, while remaining pseudomorphically strained to the GaP substrate in the $[1\bar{1}0]$ direction (see Ref.~\citenum{Ohlsson1998} and references therein). For GaAs$_{1-x}$N$_{x}$ on GaAs, the origin of the wires, also enclosed by $\{113\}$ facets and separated by several tens of nm, was similarly ascribed to strain relaxation.\cite{Suemune2000} In view of these reports, the corrugated surface morphology of the phase-separated GaP$_{1-x}$N$_{x}$ layers might be due to the relief of epitaxial strain when exceeding a critical thickness on the order of a few nm, as deduced from the 2D-3D growth mode transition observed by RHEED. However, it could also be coupled to either kinetically or thermodynamically induced lateral compositional alloy modulations\cite{Glas1987,Ipatova1994,Guyer1995,Guyer1996,Glas1997,Ipatova1998,Huang2002} as well as caused by the minimization of surface stress, which is known to result in the formation of periodically faceted surfaces even in the absence of epitaxial strain.\cite{Shchukin1999} We will return to this discussion in section~\ref{Discussion} after the complete characterization of the investigated samples.


\subsection{Microscopic analysis by SEM-EDX}


The average N incorporation detected by HRXRD was further investigated on a microscopic scale by SEM based EDX. Unlike HRXRD, which reflects the amount of N incorporated at the anion sites of the crystal lattice, EDX is sensitive to the total N concentration of the layers. Therefore, besides potentially shining light on the spatial distribution of N in phase-separated samples, EDX might also unveil the presence of significant concentrations of interstitial and antisites N atoms. The N content obtained from the EDX measurements are compiled in Table~\ref{tbl:Table1} and compared to those measured by HRXRD showing the same trend and generally a good agreement within the error margins of EDX and HRXRD measurements. As the small discrepancies between HRXRD and EDX results are not systematically biased, they cannot be unequivocally ascribed to an extra interstitial amount of N atoms detected by EDX. Observed deviations may be attributed to the measurement of different sample pieces and the smaller area over which the measurements average in the case of EDX.  Hence, down to the resolution of EDX in an SEM, on the order of a few tens of nm, the N content appears to be constant. The N segregation characterizing phase-separated samples must thus occurs on a smaller scale.

\begin{figure}
\centering
\includegraphics*[width=0.48\textwidth]{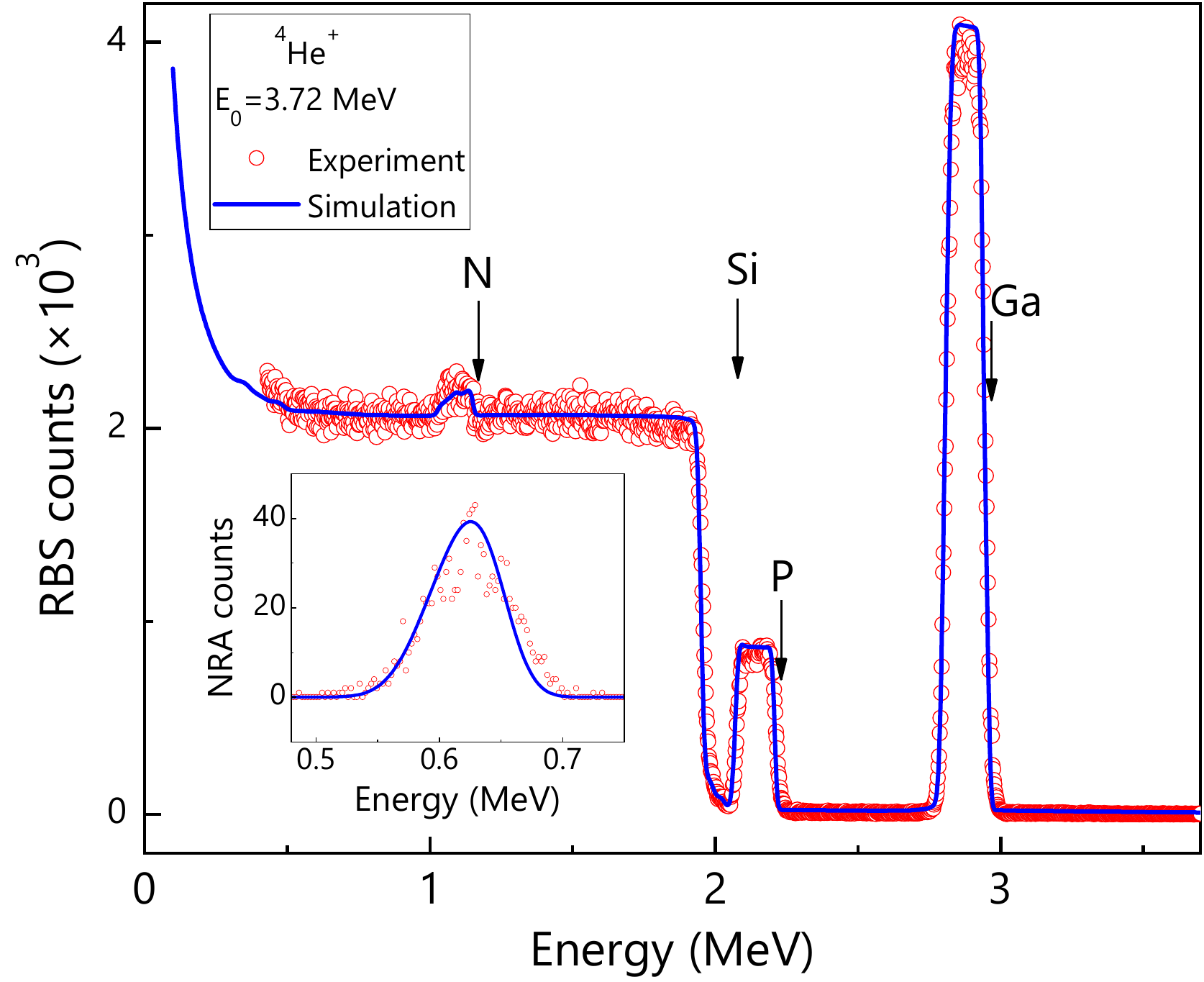}
  \caption{RBS spectrum of sample H. The inset shows the NRA spectrum of the same sample. The red circles represent the experimental data and the blue lines the simulated spectra. The arrows indicate the enegy of backscattered He$^{+}$ ions by the different elements at the sample surface. }
  \label{Figure6}
\end{figure}

\subsection{Chemical compositional analysis by RBS and NRA}
As discussed above, SEM-based EDX measurements yielded N contents rather comparable to those obtained by HRXRD, a result suggesting that there is not a remarkable concentration of non-substitutional N atoms. To further examine this matter, we analyzed by RBS and NRA the same pieces of samples C (single phase), E and H (phase separated) studied before by HRXRD in Ref.~\citenum{Saddik2021}. Notice that, similarly to EDX, both RBS and NRA are sensitive to the overall N concentration in the layers. 
 
Figure~\ref{Figure6} shows, as an example, the measured and simulated random RBS spectra of sample~H. The surface energies for Ga, P and N are indicated by arrows. Note that the low signal from N atoms, lighter than the other atoms in the layer and the substrate, overlaps with that of the Si substrate (the plateau at energies below 2~MeV), which severely limits the sensitivity of RBS to determine the N content. To overcome this problem, we acquired additional NRA spectra, as the one displayed in the inset of Fig.~\ref{Figure6} for sample~H, also including the simulated spectrum. The N signal in the NRA spectrum is clearly resolved in the 0.5 to 0.7 MeV energy range. The recorded signal is due to emitted protons, being free of backscatered alfa particles because of the mylar filter stopping power. We thus accurately determined the N content by combined simulations of RBS and NRA spectra. The alloy compositions assessed by RBS-NRA are gathered in Table~\ref{tbl:Table1}. For samples~C and E, the N contents determined by HRXRD and RBS-NRA are in good agreement within the experimental accuracy. Hence, for these two particular samples with \textit{x} values of $0.16$ and $0.28$ (HRXRD), most N atoms seems to be substitutional. In the case of sample~H, the one with the highest N concentration, the N content obtained by RBS-NRA significantly deviates from the one determined by HRXRD, namely, $0.1$ versus $0.08$. Such a deviation, consistent with the SEM-EDX measurements performed on this sample (see Table~\ref{tbl:Table1}), could be ascribed to a non-negligible fraction of N atoms incorporated at non-substitutional lattice sites or to the inaccuracy of HRXRD to measure the average N content on phase-separated samples. In general, our results compare well with the advanced channeling-RBS studies reported by Jussila et al.\cite{Jussila2014} for GaP$_{1-x}$N$_{x}$ layers grown by MOVPE on GaP$(001)$. Precisely, Jussila and co-workers found that for an \textit{x} value of $0.017$ (no far from those of samples~C and E) $91\%$ of the N atoms are substitutional, while for N mole fractions of $\approx4\%$ a large fraction incorporates at interstitial and random sites. Based on our results and those of Jussila et al., we conclude that for moderate N contents, as those required for GaP$_{1-x}$N$_{x}$ compounds lattice matched to Si ($x = 0.021$), most N atoms are incorporated in substitutional sites.


\subsection{Nanoscale analysis by TEM-related techniques}
\label{TEM section}
The analysis of the samples by SEM-EDX have shown that the compositional alloy fluctuations responsible for the two different chemical phases detected by HRXRD for samples grown within the phase-separated growth regime (see Fig.~\ref{Figure1}) must take place on a length scale below a few tens of nm. To shed light on the nature of phase separation in GaP$_{1-x}$N$_{x}$ layers grown by CBE, as well as to assess the material quality on atomic scale, single-phase and phase-separated samples were investigated in depth by TEM-related techniques.

\begin{figure}
\centering
\includegraphics*[width=0.35\textwidth]{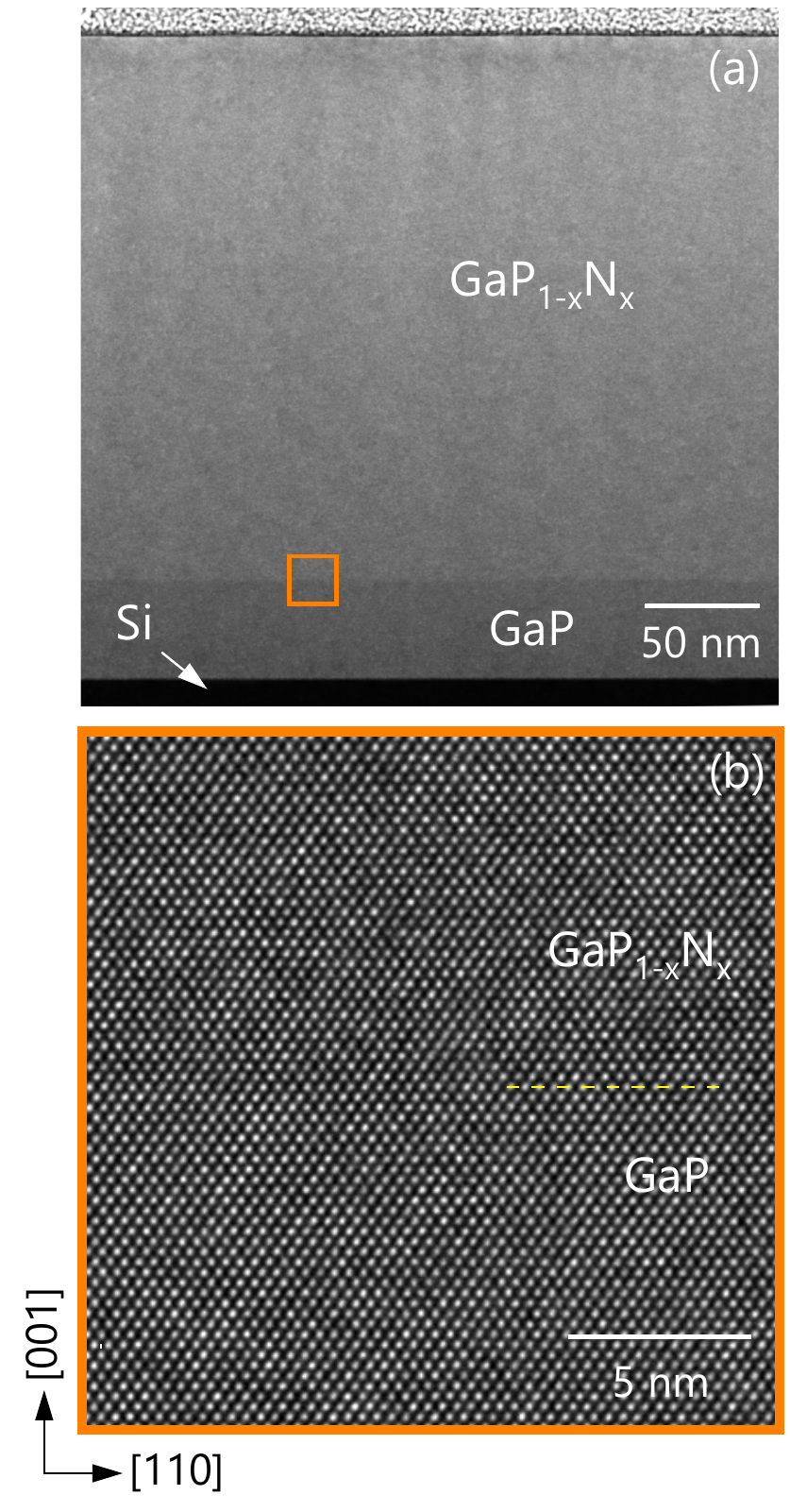}
  \caption{(a) Exemplary STEM micrograph along the zone axis $[1\bar{1}0]$ of the single-phase sample~C. (b) HRTEM micrograph of the GaP/GaP$_{1-x}$N$_{x}$ interfaces (yellow dashed line) of sample~C for the region shown by by the orange square in panel (a).}
  \label{Figure7}
\end{figure}

\begin{figure*}
\centering
\includegraphics*[width=0.65\textwidth]{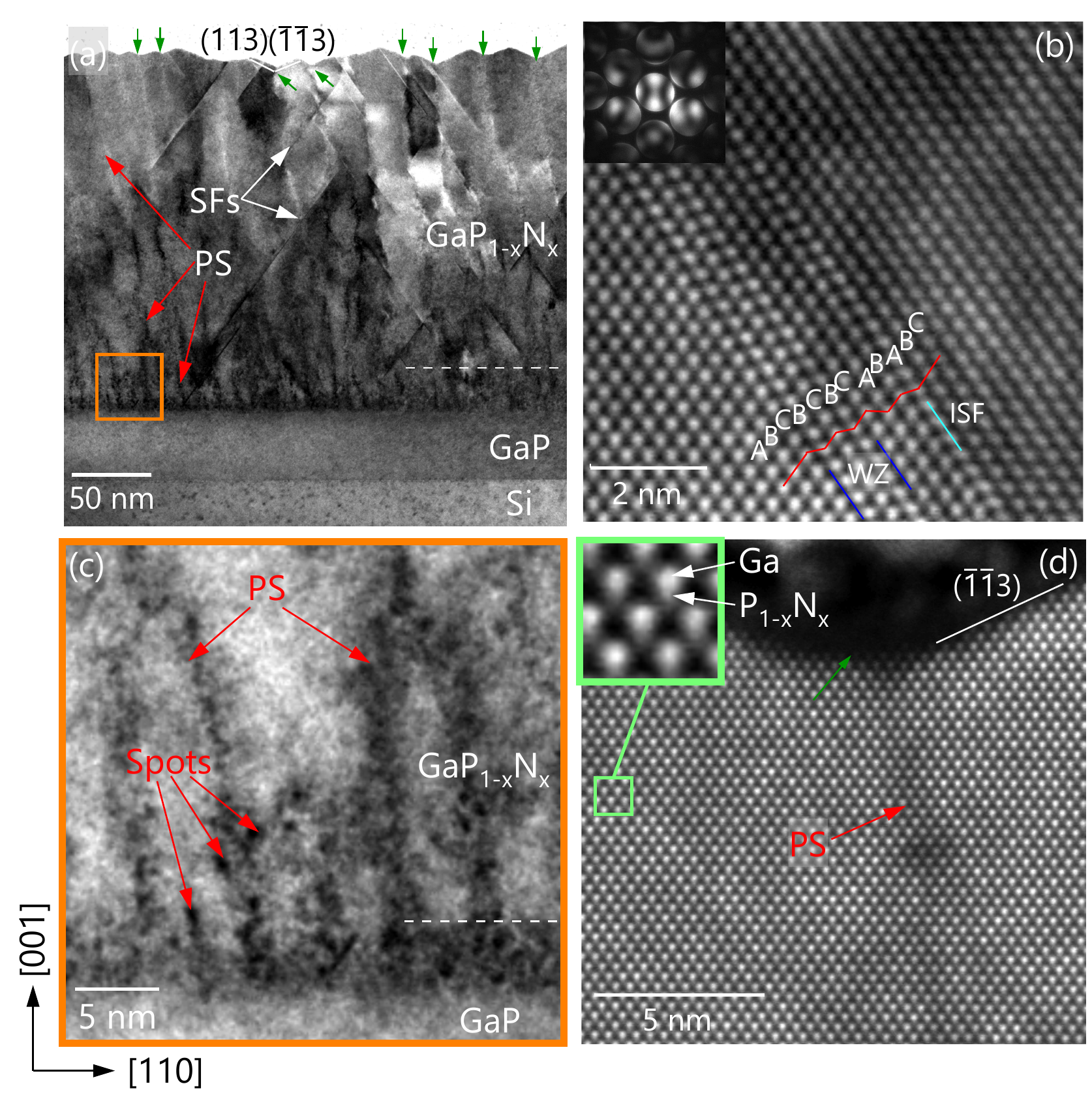}
  \caption{(a) Exemplary BF micrograph along the zone axis $[1\bar{1}0]$ of the phase-separated sample~G. The dashed line roughly indicates a boundary of the layer with an almost periodic distribution of regions with phase separation (PS). The green arrows note troughs. (b) HRSTEM image illustrating an atomic structure of SFs in the latter sample. The light blue line is an intrinsic stacking fault. The region with a wurtzite structure is bounded by the dark blue lines. The CBED pattern shown in the inset demonstrates $[1\bar{1}0]$ zone axis orientation. (c) Magnified BF micrograph of the region indicated by the orange square in panel (a). The dashed line roughly indicates a boundary of layer almost without PS and higher nitrogen content. (d) HRSTEM image illustrating an atomic structure of the material near a trough noted by the green arrow. The inset shows the magnified image of the region and confirms $[1\bar{1}0]$ zone axis orientation.}
  \label{Figure8}
\end{figure*}

Figure~\ref{Figure7}(a) presents a STEM image of a single-phase sample, namely, sample~C. In this micrograph, the region with the slightly higher intensity corresponds to the GaP$_{1-x}$N$_{x}$ layer, despite the fact that some of the P atoms are replaced by lighter N atoms. The origin of the contrast with respect to the GaP buffer layer can be explained by the tetragonal distortion of the crystal lattices. Since the $[110]$ planes of both layers are fitted to the Si substrate and the lattice parameter of the GaP$_{1-x}$N$_{x}$ layer is smaller than that of GaP,\cite{BenSaddik2019} interplanar distances between $[001]$ planes are smaller in the GaP$_{1-x}$N$_{x}$ layer. As a result, the number of atomic planes per unit length along the $[001]$ direction for the GaP$_{1-x}$N$_{x}$ layer is larger than for the GaP buffer layer located below. Apart from the different brightness of the GaP$_{1-x}$N$_{x}$ and GaP layers, their homogeneous intensity distribution indicates the absence of extended defects and of local compositional variations in the case of GaP$_{1-x}$N$_{x}$. The high structural quality of both layers and the GaP$_{1-x}$N$_{x}$/GaP interface is additionally confirmed by the HRTEM image shown in Fig.~\ref{Figure7}(b).

\begin{figure}
\centering
\includegraphics*[width=0.35\textwidth]{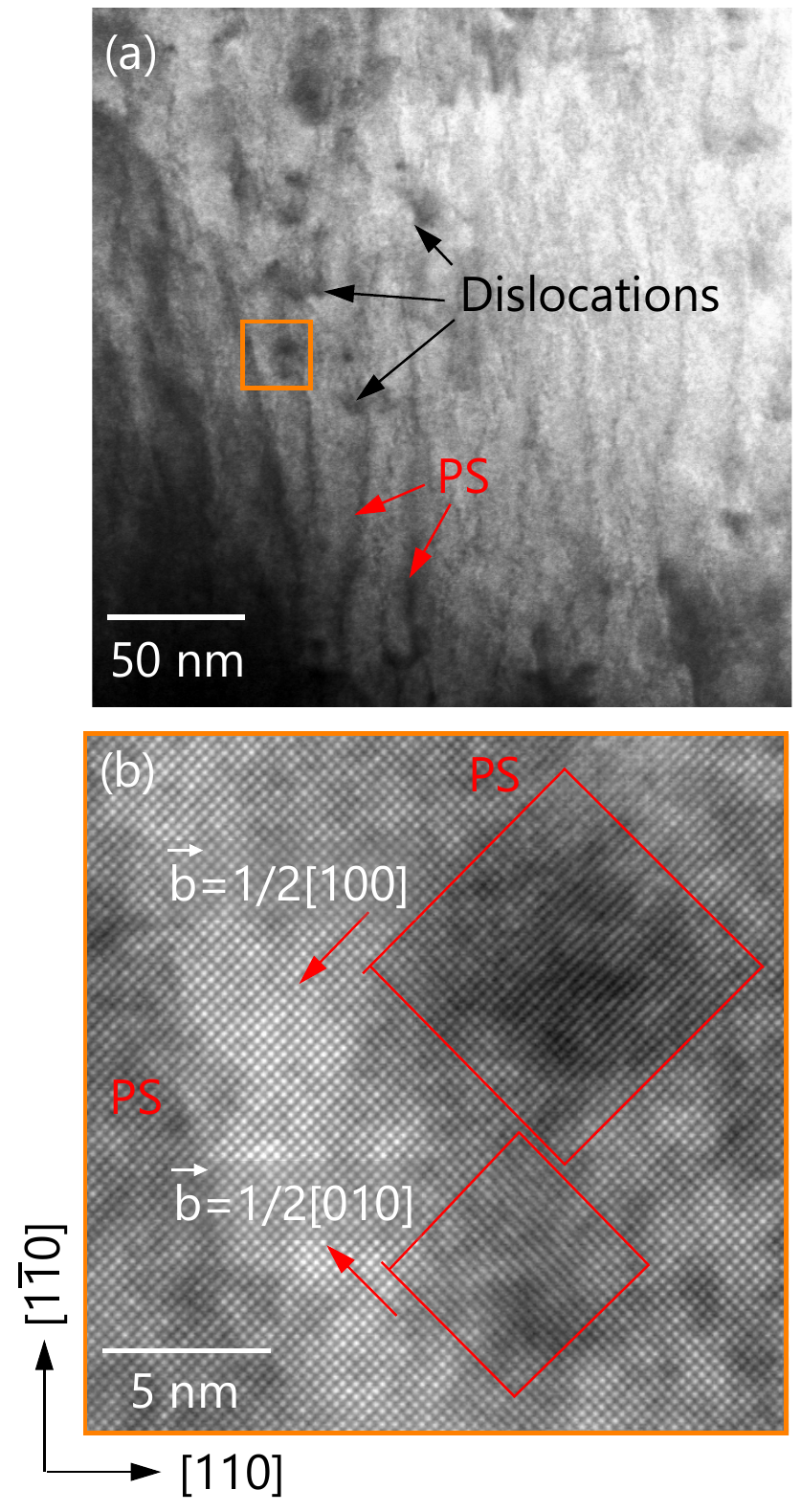}
  \caption{(a) Plan-view STEM micrograph of the phase-separated sample~G. (b) HRSTEM image illustrating an atomic structure of the spots indicated by the orange square in panel (a). The red contours are open Burger’s circuits corresponding to marked projections of Burger’s vectors of dislocations.}
  \label{Figure9}
\end{figure}

\begin{figure*}
\centering
\includegraphics*[width=0.7\textwidth]{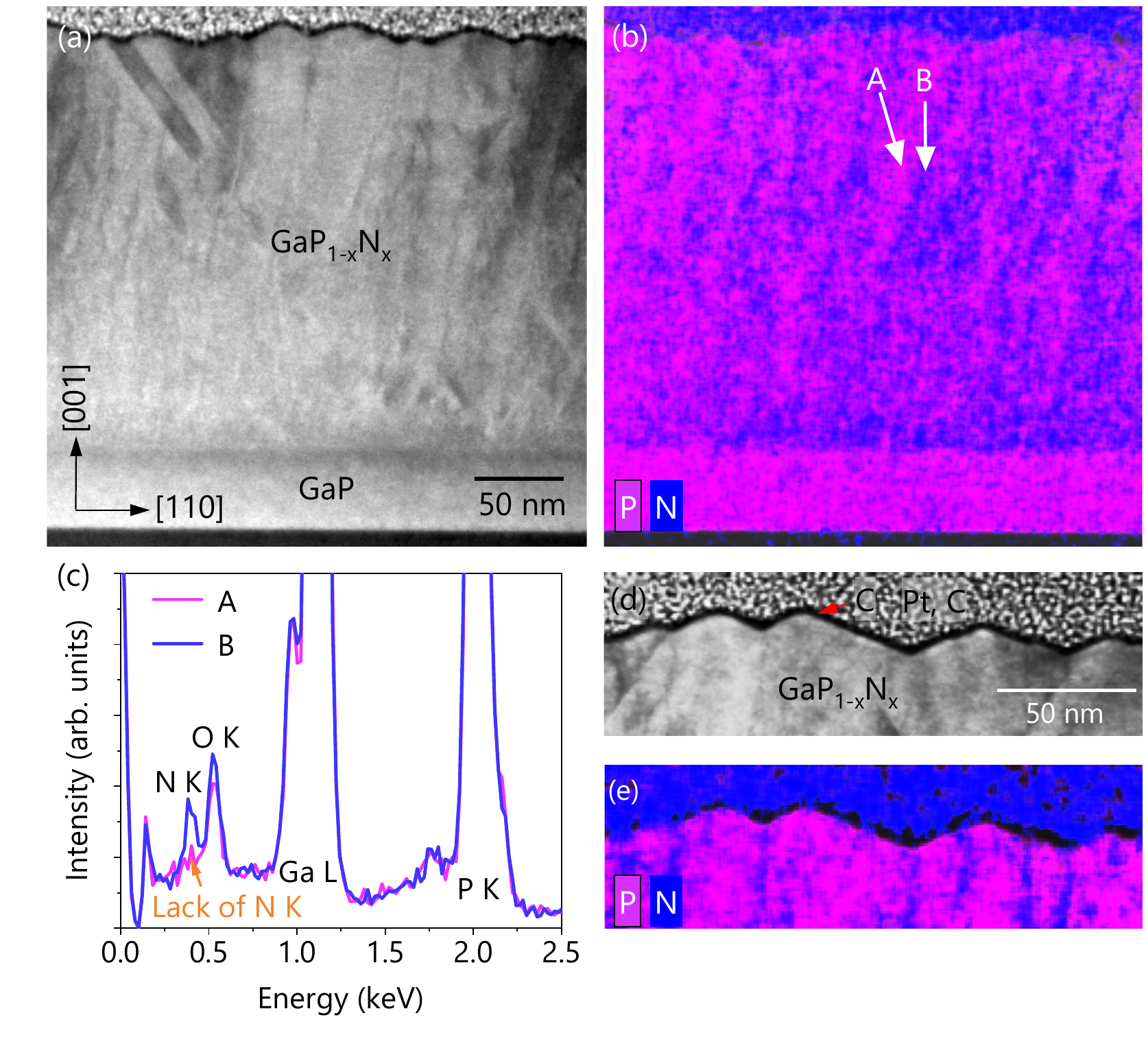}
  \caption{(a) STEM image along the zone axis $[1\bar{1}0]$ and (b) corresponding EDX map illustrating the N and P distribution within sample~G. (c) EDX spectra registered from the regions indicated by letters A and B in (b). (d) Magnified STEM image along the zone axis $[1\bar{1}0]$ and (b) corresponding EDX map illustrating the N and P distribution near surface of sample~G.}
  \label{Figure10}
\end{figure*}

Unlike sample~C, the crystal structure of the chemically phase-separated samples includes several types of defects, as can be seen in Fig.~\ref{Figure8}(a) for sample~G. The inclined sharp lines observed in this BF micrograph are due to stacking faults (SFs), which often appear in multilayer structures with misfit of crystal planes.\cite{Ding2020,Gundimeda2022} The presence of SFs in these regions was actually evidenced by HRSTEM, as shown in Fig.~\ref{Figure8}(b), where an intrinsic stacking fault (ISF) is observed (indicated by the light-blue line). The analysis of numerous micrographs allowed us to conclude that ISFs are the most abundant extended defects at inclined sharp lines as those seen in Fig.~\ref{Figure8}(a). Besides ISFs, less often we found extrinsic stacking faults (not shown here) and thin strips with a wurtzite-like structure (WZ), as the region in between the dark-blue lines in Fig.~\ref{Figure8}(b).

Returning to Fig. 8(a), the strips imaged in BF-micrographs as rather narrow and elongated bright or dark regions oriented almost parallel to the $[001]$ direction and indicated by red arrows are caused by variations in the chemical composition due to chemical phase separation (PS), as we will demonstrate in the following. Such variations give rise to local tetragonal distortions of the crystal lattice changing the condition of electron diffraction in both the PS strips and the surrounding material. These PS strips are nearly periodically distributed along the $[110]$ direction with a mean separation distance of about $10$~nm in the vicinity of the GaP/GaP$_{1-x}$N$_{x}$ interface. Interestingly, at a distance of approximately $40$~nm from the interface, as marked by the dashed line inserted in Fig.\ref{Figure8}(a), the PS strips begin to join together, and therefore, their lateral density decreases. Additional important details related to the chemical phase-separation process can be observed in Fig.\ref{Figure8}(c), where we present a magnified micrograph of the region near the GaP/GaP$_{1-x}$N$_{x}$ interface. In the first $5$~nm of the GaP$_{1-x}$N$_{x}$ layer, the image intensity is lower and more homegeneous than in the surrounding regions, and the PS strips are much less pronounced (in case they exist). This darker region is caused by a higher N concentration, implying an enhanced and more homogeneous incorporation of N at the beginning of the GaP$_{1-x}$N$_{x}$ layer. We also find in Fig.~\ref{Figure8}(c) dark spots associated to crystal lattice distortions within both the PS vertical wires and the $5$~nm thick darker region formed at the onset of the GaP$_{1-x}$N$_{x}$ layer. An additional examination of the dark spots by HRTEM did not show dislocations there (not shown here). Hence, these objects probably correspond to N-rich nanoclusters inside the main phase, but this assumption should be verified by additional studies.

With regards to the surface morphology of sample~G, the micrographs shown in Figs.~\ref{Figure8}(a) and \ref{Figure8}(d) evidence the pronounced roughness observed by AFM [Fig.~\ref{Figure4}(c)]. In agreement with the $[1\bar{1}0]$ RHEED patterns of phase-separated samples, the images show that the surface is mainly composed of $(113)$ and $(\bar{1}\bar{1}3)$ facets. At the joint between these facets, we observe the formation of peaks and troughs [the latter are noted by green arrows in Figs.\ref{Figure8}(a) and \ref{Figure8}(d)]. According to further micrographs, SFs leave the crystal at random positions, so they have no connection with surface irregularities. In contrast, the periodically faceted surface correlates with PS regions. Specifically, at the troughs we always detect PS wires, where the lower intensity of the atomic columns with respect to the surrounding regions in HRSTEM micrographs [see Fig.\ref{Figure8}(d)] reveals a lower mean atomic number of electron scattering atomic columns in them. In consequence, the N/P substitutional ratio is higher at the troughs. Based on this observation, we conclude that the periodically faceted surface morphology is intimately correlated with the lateral compositional fluctuations, as theoretically proposed and experimentally observed in other III-V compounds such as GaAs$_{1-x}$N$_{x}$, In$_{1-x}$Ga$_{x}$P, and GaAs$_{1-x}$Bi$_{x}$.\cite{Ipatova1994,Guyer1995,Guyer1996,Glas1997,Ipatova1998,Huang2002,Suemune2000,Bortoleto2007,Tait2018,Luna2019} 

Sample~G was also imaged by plan-view STEM. The results, illustrated in Fig.~\ref{Figure9}(a), confirm the structure seen by AFM. Here, we see that the PS regions, recognizable as dark strips, are slightly uneven and elongated along the $[1\bar{1}0]$ direction. Also, the image contains dark spots noted as dislocations, which are usually located at the bend points of strips. Figure~\ref{Figure9}(b) shows a HRSTEM image of some the spots seen in Fig.~\ref{Figure9}(a). The spatial resolution of this micrograph makes it possible to visualize Ga atomic columns, while those of P and N are not resolved. Burger’s circuits built around dark areas, which corresponds to spots in Fig.\ref{Figure9}(a), are open indicating the presence of dislocations. Their mismatch vectors are equal to $1/2[100]$ and $1/2[010]$ and represent the projections of Burgers vectors from dislocations on the $(001)$ plane. 

To gain additional insights into the chemical fluctuations of sample~G, it was also studied by TEM-EDX. Figures~\ref{Figure10}(a) and \ref{Figure10}(b) present a STEM image and its corresponding EDX map, respectively. The STEM micrograph reveals the same structure as the BF image shown in Fig.\ref{Figure8}(a). With respect to the EDX map, it shows the superimposed X-ray radiation signals from P and N atoms. Ga and O signals were also detected, but they are not shown here because Ga is homogeneously distributed and O is an artifact caused by the exposure of the specimen to the air. Though the shown EDX map is qualitative owing to the low concentration of N atoms and does not permit to conclusively discern phase-separated regions at the bottom part of the GaP$_{1-x}$N$_{x}$ layer, it is enough to draw some important conclusions as further discussed here. At the middle and top parts of the GaP$_{1-x}$N$_{x}$ layer, we clearly distinguish pink areas and blue stripes corresponding to regions with comparatively low and high N concentrations, respectively. The direct comparison of the EDX map with its corresponding STEM image demonstrates that the nearly vertical strips seen in the micrograph are associated to regions with higher N concentrations, as already inferred from the analysis of the images shown in Figs.~\ref{Figure8}(a) and \ref{Figure8}(c). The higher concentration of N at those areas is further illustrated in Fig.~\ref{Figure10}(c), where we plot two different EDX spectra acquired at the specimen regions marked in Fig.~\ref{Figure10}(b) as A and B, corresponding to pink and blue ares of the EDX map, respectively. It is clear that while the spectrum acquired at position B exhibits a N-related line, the one taken at position A lacks that peak. Hence, we conclude that the N concentration in the pink areas of the EDX map is below $1-2\%$, as estimated from the low-counts in the spectrum. Last, we made a closer examination of the N distribution near the sample surface. The results are summarized in the highly-magnified HAADF micrograph shown in Fig.~\ref{Figure10}(d) and the corresponding EDX map shown in Fig.~\ref{Figure10}(e). These images evidence a higher incorporation of N at all the troughs formed between the inclined $(113)$ and $(\bar{1}\bar{1}3)$ facets. 

A quantitative EDX analysis was also performed for samples~D, F and G. Sample C was also investigated by EDX, but its composition analysis was impossible due to the very low intensity of the N X-ray peak associated to its low N content. Table~\ref{tbl:Table1} summarizes, first, the composition averaged over the whole thickness of the GaP$_{1-x}$N$_{x}$ layer for samples D--G, and second, the compositions of the bottom and top-half part of the GaP$_{1-x}$N$_{x}$ layer for samples F and G. The composition averaged over the whole layer thickness for samples~D--G are in reasonable agreement within the experimental error with the results obtained by HRXRD. Instead, SEM-EDX should be compared to TEM-EDX in the upper half-part of the layer, as the former technique primarily probes the sample surface; both results also show a good agreement. The analysis of the composition near the GaP/GaP$_{1-x}$N$_{x}$ interface for samples F and G evidences an improved N incorporation efficiency before the lateral composition fluctuations set in. According to the EDX study, we conclude that: (i) the mean N content at the bottom part of GaP$_{1-x}$N$_{x}$ phase-separated layers is more homogeneous and higher than at the middle and upper parts, where phase separation becomes very pronounced, (ii) the regions where nearly vertical wires are observed in BF and STEM micrographs indeed correspond to N-rich areas, and (iii) the N-rich regions are correlated with the periodically faceted surface since N atoms are seen to be preferentially incorporated at the troughs formed between $(113)$ and $(\bar{1}\bar{1}3)$ facets.

\section{\label{Discussion}Discussion}
\begin{figure}
\includegraphics*[width=0.5\textwidth]{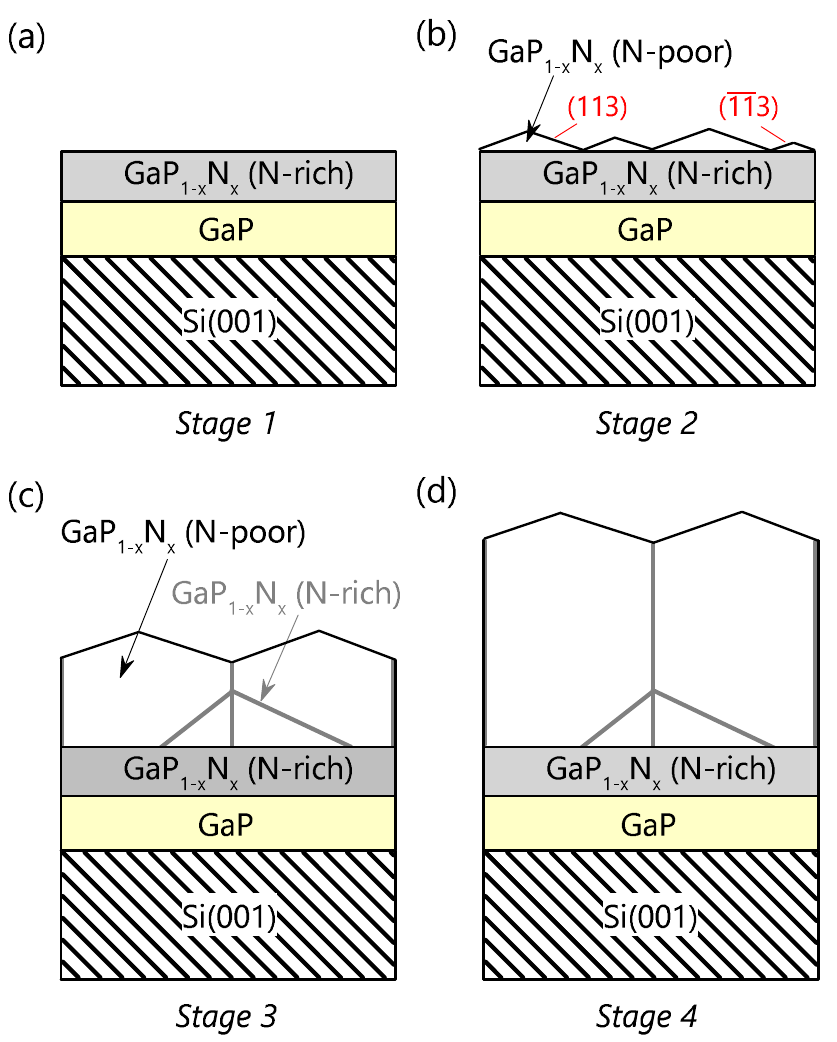}
\caption{Schematic illustration depicting the time evolution of phase-separated GaP$_{1-x}$N$_{x}$ layers through four different stages: (a) layer-by-layer growth with homogeneous chemical composition, (b) 2D--3D growth mode transition leading to onset of phase separation as result of $\{113\}$ facets formation, (c) merging of N-rich regions, and (d) steady-state growth of phase-separated layer. For clarity, the thickness on the substrate and the layers are not to scale.}
\label{Figure11}
\end{figure}

As follows from the studies reported here as well as in Ref.\citenum{Saddik2021}, single-phase GaP$_{1-x}$N$_{x}$ compounds can be grown layer-by-layer by CBE with a high structural quality up to a N mole fraction of about $4\%$. This type of samples are free of SFs and characterized by isotropic surface morphologies with rms roughness values below $0.4$~nm. Therefore, from growth and structural points of views, the synthesis of GaP$_{1-x}$N$_{x}$ compounds lattice-matched to Si ($x = 0.021$) is totally feasible. 

As the growth temperature is decreased and/or the flux of DMHy is raised to increase the N mole fraction, at a certain point the samples become chemically phase separated. As can be seen in the growth diagram shown in Fig.~\ref{Figure1}, the maximum N mole fraction that can be incorporated before chemical phase separation sets in depends on the specific growth conditions. In particular, the use of lower substrate temperatures helps to delay the onset of phase separation as the N content is increased by raising the flux of DMHy. Chemical phase separation comes together with the formation both extended defects (SFs and dislocations) and a corrugated surface morphology consisting of wires enclosed by $\{113\}$ facets and aligned along the $[1\bar{1}0]$ with a period of several tens of nm. The segregation of the GaP$_{1-x}$N$_{x}$ layer in regions with two distinct N contents is  intimately related to the corrugated surface morphology, as demonstrated here by the lateral compositional modulations observed by TEM along the $[110]$ direction, where a higher N mole fraction is detected at the troughs formed in between wires. Hence, the incorporation efficiency of N atoms at the troughs must be enhanced with respect to the flat $\{113\}$ facets. Interestingly, this N distribution is opposite to that reported in Ref.~\cite{Suemune2000} for GaAs$_{1-x}$N$_{x}$ layers grown on GaAs by CBE. In this latter case, N was found to incorporate less efficiently at the troughs formed at the joints between $\{113\}$ facets in phase-separated GaAs$_{1-x}$N$_{x}$ layers with an identically corrugated surface morphology, a result explained by Suemune and co-workers in terms of a preferential incorporation of N into A(Ga)-steps on $\{11n\}$ surfaces.\cite{Suemune2000} 

According to the analysis of our samples, the growth of phase-separated GaP$_{1-x}$N$_{x}$ layers proceeds through four different growth stages, as schematically illustrated in Fig.~\ref{Figure11} and further described below: 

(i) \textit{Stage 1: Layer-by-layer growth with homogeneous chemical composition} [Fig.~\ref{Figure11}(a)]. As demonstrated by the observation of RHEED intensity oscillations [Fig.~\ref{Figure3}(a)], the GaP$_{1-x}$N$_{x}$ layers initially grow layer-by-layer up to a thickness of at least $2-3$~nm. Even though GaP$_{1-x}$N$_{x}$ grows as a conglomeration of sub-nanometer clusters, at this stage the chemical composition is homogeneous on a microscope scale. 

(ii) \textit{Stage 2: 2D--3D growth mode transition, the onset of phase separation} [Fig.~\ref{Figure11}(b)]. Upon reaching a certain critical thickness, there is a 2D--3D growth mode transition associated to a decrease in the mean N content as well as to the formation of $\{113\}$ faceted wires along the $[1\bar{1}0]$ direction and the creation of lateral compositional modulations. We have determined the experimental critical thickness by TEM for two samples, F and G. The corresponding critical thicknesses are $8$~nm for sample~F and $5$~nm for sample~G, which are compatible with the observation of RHEED intensity oscillations up to a layer thickness of about 3.3 and 2.2~nm, respectively. These critical thicknesses are too small for the fabrication of thick-film based devices, but enough to fabricate the active region of optoelectronic devices based on quantum heterostructures. 

(iii) \textit{Stage 3: Merging of N-rich regions} [Fig.~\ref{Figure11}(c)]. As the growth proceeds, the wires widen and some of them coalesce with each other resulting in the formation of thicker wires. This scenario is supported by the TEM micrographs of sample~G [see Fig.~\ref{Figure8}(a) and \ref{Figure8}(c)], where we directly observe a varying period of the lateral compositional fluctuations along the growth direction as well as the merging of N-rich regions for a layer thickness of 20--40~nm. The decreasing lateral density of troughs associated to the merging of wires is likely caused by growth rate fluctuations among adjacent $\{113\}$ facets.\cite{Spencer1992,Guyer1995} 

(iv) \textit{Stage 4: Steady-state growth} [Fig.~\ref{Figure11}(d)]. Upon the partial merging of N-rich regions, the growth reaches steady-state conditions, i.\,e., the coupled morphological-compositional fluctuations reach a well defined period on the order of a few tens of nm along the $[110]$ direction. Interestingly, the filamentary like N-rich regions ending at the troughs formed in between $(113)$ and $(\bar{1}\bar{1}3)$ facets are not perfectly perpendicular to the substrate surface (as shown in the sketch), but exhibit a tilt of $5-8$~$^{\circ}$ with respect to the substrate surface normal, as seen in some of TEM micrographs acquired along the $[1\bar{1}0]$ zone axis [see Figs.\ref{Figure8}(a) and \ref{Figure10}(a)]. Hence, the positions of the troughs shifts during growth along the $[110]$ direction as theoretically discussed in Refs.~\citenum{Spencer1992} and \citenum{Guyer1995}.

The formation of coupled lateral compositional and morphological modulations have been reported for several II-VI and III-V alloys,\cite{Henoc1982,Ueda1984,Mahajan1984,Norman1985,Chu1985,Ueda1988,Ueda1989,UEDA1989_1,McDevitt1992,Kuo1994,Jun1996,Suemune2000,Tait2018,Luna2019} but not for GaP$_{1-x}$N$_{x}$ compounds. Lateral compositional and morphological modulations can be caused by kinetic and thermodynamic phenomena which, for a given material, depend on multiple factors such as diffusion barriers, epitaxial strain, substrate temperature and growth rate.\cite{Glas1987,Ipatova1994,Guyer1995,Guyer1996,Glas1997,Ipatova1998,Huang2002} From a kinetic point of view, the planar and chemically homogeneous surface of an alloy can be destabilized if it is composed of atoms with large atomic-size differences,\cite{Guyer1996,Ipatova1998} as in the case of GaP$_{1-x}$N$_{x}$. The instability arises, on the one hand, from the random alloy fluctuations created at the topmost layer as the atoms are buried while migrating on the surface. As a result of the composition-dependent lattice parameter, the random alloy fluctuations formed at the topmost layers create a strain field at the surface that influences adatoms migration, which is then not exclusively controlled by surface diffusion. There is thus a strain-induced drift that favors the incorporation of the biggest atoms at the regions with a larger local lattice parameter and vice-versa, yielding to the formation of compositional and morphological irregularities.\cite{Guyer1996,Ipatova1998} On the other hand, for a coherently strained film, a morphological instability can arise due to elastic strain relaxation. For example, as described by Guyer and Voorhees, the stress of a film with a sinusoidally perturbed surface is relaxed at the wave crests and increased at the troughs.\cite{Guyer1996} This phenomenon also causes a strain field at the surface that influences the local chemical composition due to again strain-induced drift of surface adatoms. The creation of coupled morphological-compositional modulations may also have instead a purely thermodynamic origin as described by Glas.\cite{Glas1997} The theoretical work of Glas shows that any nonhydrostatically stressed alloy with a free surface and a stress-free unit cell whose volume depends on composition is unstable against to a range of coupled surface undulations and composition modulations. The coupled morphological-compositional modulations, whose occurrence only requires an atomic-size difference between the alloy constituents, is driven by the reduction of the elastic energy with respect to either the case of pure surface undulation or that of pure composition modulation. Unstable coupled morphological-compositional modulations exist at any temperature, so that the overall critical temperature above which the alloy is chemically homogeneous and flat is infinite. These alloys should thus tend to decompose at any temperature in the presence of adequate mass-transport mechanisms, i.\,e., provided that the process is not kinetically impeded. Nevertheless, as explicitly discussed by Glas, his calculations do not explain how an initially chemically homogeneous and flat alloy evolves to reach thermodynamic equilibrium, i.\,e., until forming a layer with coupled morphological-compositional modulations. This is instead a kinetically governed process, as shown for instance by Huang and Desai.\cite{Huang2002} According to Refs.~\citenum{Guyer1996} and \citenum{Huang2002}, the critical thickness above which coupled morphological-compositional fluctuations appears is proportional to the ratio between the deposition and perturbation growth rates. Even though in our case we cannot conclude on whether the underlying mechanism for the observed growth instabilities is primarily kinetically or thermodinamically driven, the kinetic nature of the critical thickness implies that by increasing the deposition rate, a parameter not explored in the present work, it might be possible to increase the layer thickness of single-phase GaP$_{1-x}$N$_{x}$ compounds grown by CBE with $x>0.04$ beyond the values found in this study.

\section{Summary and conclusions}
We have analyzed the growth mode, the chemical composition and the structural quality of GaP$_{1-x}$N$_{x}$ layers grown by CBE on GaP/Si(001) substrates as a function of the growth temperature and the flux of the N precursor. For N mole fractions below $0.04$, it is possible to obtain single-phase layers with a homogeneous chemical composition, a high structural quality, and a smooth surface morphology. This type of samples, easier to synthesize at comparatively lower growth temperatures, were found to grow according to the Frank-Van der Merwe (layer-by-layer) growth mode. As the N content is increased, by either lowering the growth temperature or increasing the flux of the N precursor, the GaP$_{1-x}$N$_{x}$ quality eventually degrades resulting in chemically phase-separated layers with a poor morphological and structural quality. These layers, besides containing SFs and dislocations, are characterized by periodic nanoscopic lateral compositional fluctuations coupled to morphological irregularities along the $[110]$ direction. In particular, N incorporation is found to be enhanced at troughs created at the joints between periodically arranged $\{113\}$ facets. Importantly, despite of the final 3D surface morphology, the growth initially proceeds in a layer-by-layer fashion with a homogeneous chemical composition up to reaching a critical thickness of several nm, as demonstrated by the observation of RHEED intensity oscillations and the analysis of the samples by TEM, respectively.  

On the basis of our study, we conclude that: (i) CBE is suitable for the fabrication of at least several hundreds of nm thick GaP$_{1-x}$N$_{x}$ layers with a homogeneous chemical composition up to a N mole fraction of 0.04; a value well above the one required for the lattice-matched integration of Ga$_{1-x}$N$_{x}$-based devices on Si, and (ii) N mole fractions above 0.04 could be achieved while preserving a high structural quality provided that the layer thickness does not exceed the critical one for the 2D--3D growth mode transition, which might be increased with respect to the values found in this work ($5-8$~nm) by increasing the deposition rate. Nevertheless, since the experimentally observed critical thicknesses are on the order of a few nm, it is still possible to use GaP$_{1-x}$N$_{x}$ compounds with $x>0.04$ in quantum heterostructures such as quantum wells and the short-period superlattices theoretically proposed by Kharel and Freundlich in Ref.~\citenum{Kharel2018} to pseudomorphically integrate III-V solar cells on Si.

\section*{Data Availability Statement}

The data that support the findings of this study are available
from the corresponding author upon reasonable request.

\begin{acknowledgments}
We thank Dr. Jonas Lähnemann for his invaluable assistance in carrying out the EDX measurements in SEM. We are indebted to Dr. Esperanza Luna from Paul-Drude-Institut because of in depth discussions on chemical phase-separation in highly mismatched alloys. We also thank to T. Vallés for his technical support and assistance. This work was supported by the former Ministerio de Ciencia, Innovación y Universidades (Project No. TEC2016-78433-R), the current Ministerio de Ciencia e Innovación (Project No. PID2020-114280RB-I00), and the Ministry of Science and Higher Education of the Russian Federation (State assignment No. FSMR-2020-0018). Additionally, S. Fernández-Garrido acknowledges the financial support received through the program Ramón y Cajal (co-financed by the European Social Fund) under Grant No. RYC-2016-19509 from Ministerio de Ciencia, Innovación y Universidades. We also acknowledge the service from the MiNa Laboratory at IMN, and funding from CM (project S2018/NMT-4291 TEC2SPACE), MINECO (project CSIC13-4E-1794) and EU (FEDER, FSE).
\end{acknowledgments}

\section*{References}
\nocite{*}
\bibliographystyle{plain}

\providecommand{\noopsort}[1]{}\providecommand{\singleletter}[1]{#1}%

\end{document}